\documentclass[12p]{article}
\usepackage{latexsym}
\usepackage{color}
\usepackage{amsmath}
\usepackage{epsfig}
\usepackage{hyperref}
 \usepackage{subfigure}
 \usepackage{dcolumn}
   \usepackage{threeparttable}

\newcommand{\ortala}[1]{\begin{center}#1\end{center}}

\newcommand{\sandd}[1]{\left\langle #1\right\rangle}

\newcommand{\summ}[3]{{{\underset{#1 }{\overset{#2}{\displaystyle\sum}}}#3}}

\newcommand{\re}[1]{(\ref{#1})}

\newcommand{\eq}[2]{\begin{equation}\label{#1}  #2\end{equation}}

\newcommand{\paran}[1]{\left(#1\right)}
\newcommand{\parank}[1]{\left[#1\right]}

\newcommand{\sch}[1]{Schrodinger}

\newcommand{\komb}[2]{\paran{\begin{array}{c} #1 \\ #2 \end{array}}}

\usepackage{amssymb}
\usepackage{latexsym}
\begin{document}

\ortala{\large\textbf{Anisotropic Heisenberg Model in Thin Film Geometry}}

\ortala{\textbf{\"Umit Ak\i nc\i \footnote{\textbf{umit.akinci@deu.edu.tr}}}}

\ortala{\textit{Department of Physics, Dokuz Eyl\"ul University,
TR-35160 Izmir, Turkey}}

\section{Abstract}

The effect of the anisotropy in the exchange interaction on the phase diagrams and magnetization behavior of the  Heisenberg thin film has been investigated with effective field formulation in a two spin cluster using the decoupling approximation. Phase diagrams and magnetization behaviors have been obtained for several different cases, by grouping the systems in accordance with, whether the surfaces/interior of the film have anisotropic exchange interaction or not. Particular attention has been paid on the evolution of the special point coordinate with the anisotropy in the exchange interaction. It has been concluded that for increasing ratio of the anisotropies in the exchange interaction between the surface and interior of the film, special point disappears at a special value of this ratio.

\section{Introduction}\label{introduction}

Recently, there has been growing interest both theoretically and
experimentally on the finite magnetic materials especially on semi-infinite systems and thin films.
The magnetic properties of the materials in the presence of the free surfaces are drastically different from the bulk counterparts. This is because of the fact that, free surface breaks the translational symmetry, i.e. surface atoms are embedded in an environment of lower symmetry than that of the inner atoms \cite{ref1,ref2}. If the strength of the surface exchange interaction is greater than a critical value, the surface region can exhibit an ordered phase even if the bulk is paramagnetic and it has a transition temperature higher than the bulk one. This fact has been observed experimentally\cite{ref3,ref4,ref5}.

The development of the molecular beam epitaxy technique and its application to the growth of thin metallic films has stimulated renewed interest in thin film magnetism. In a thin film geometry, one important  question is dependence of the thermodynamic properties and Curie temperature of the film on the thickness of the film. It was experimentally found that, the Curie temperature and the average magnetic moment per atom increases with the increasing thickness of the film \cite{ref6,ref7}. This thickness dependent Curie temperature also has been measured in Co\cite{ref8}, Fe\cite{ref9} and Ni\cite{ref10} films.

In order to get insight on the phase transition characteristics and thermodynamic properties of thin films, it is valuable in theoretical manner to solve related models in thin film geometry such as Hesinberg and Ising models. One class of the films
which exhibit a strong uniaxial anisotropy\cite{ref11} can be modeled by Ising model. These systems have been widely studied in literature by means of several theoretical methods such as Monte Carlo (MC) simulations \cite{ref12},  mean field approximation (MFA) \cite{ref13} and effective field theory (EFT) \cite{ref14}. In the Ising case, the study of ferroelectric films treated by transverse Ising model (TIM) are more common in the literature since fabricated ferroelectric thin films such as Barium titanate $BaTiO_3$ and Strontium titanate  $SrTiO_3$ \cite{ref15,ref16,ref17} have a lot of technological applications in optoelectronics \cite{ref18} and microelectronics \cite{ref19}. The TIM on the thin film geometry can be solved by a variety of techniques such as MFA \cite{ref20} and EFT\cite{ref21}. Besides, within the EFT formulation,  the effect of the dilution on the phase diagrams and magnetic properties of the ferroelectric thin films (by means of the TIM) has been studied\cite{ref22} and  also long range interactions have been taken into account \cite{ref23}. In order to simulate more realistic models, different exchange interactions have been defined and different transverse fields have been considered on the surface and bulk, which anticipates the mimics of the surface effects within the framework of EFT \cite{ref24}. Another improvement of the Ising thin film models is the consideration of amorphisation of the surface  due to environmental effects, and these realizations have also been solved using EFT \cite{ref25}. There are also higher spin Ising thin films, e.g. spin-1 Ising thin films have been studied \cite{ref26,ref27}.

Although the Ising model has been studied on the thin film geometry widely, less attention was paid on the Heisenberg model on thin film geometry, in comparison with the Ising model. Thin films which do not exhibit a strong uniaxial anisotropy requires to solve the Heisenberg model in the thin film geometry. In order to see the effect of the presence of the surface in the system on the critical and thermodynamical properties, Heisenberg model on a semi infinite geometry has been solved using a wide variety of techniques such as Green function method \cite{ref28}, renormalization group technique \cite{ref29}, MFA \cite{ref30}, EFT \cite{ref31,ref32,ref33},
high temperature series expansion \cite{ref34,ref35}. As in the case of Ising counterpart, Heisenberg model in a thin film geometry has been solved in a limited case. For instance, Heisenberg model on a thin film geometry with Green function method \cite{ref36,ref37,ref38}, renormalization group technique \cite{ref39}, MFA \cite{ref30}, EFT \cite{ref40_ek,ref40} and
MC \cite{ref41,ref42}, are among them.

The aim of this work is the determine the effect of the anisotropy in the exchange interaction on the phase diagrams and magnetization behavior of the  Heisenberg thin film. For this aim, the paper is organized as follows: In Sec. \ref{formulation} we
briefly present the model and  formulation. The results and
discussions are presented in Sec. \ref{results}, and finally Sec.
\ref{conclusion} contains our conclusions.

\section{Model and Formulation}\label{formulation}

Thin film can be treated as a layered structure which consist of interacting $L$ parallel layers. Each layer is defined as a regular lattice with coordination number $z$. When we choose $z=4$, this means that each layer has square lattices and each nearest neighbor layer have interaction.
The Hamiltonian of the thin film is given by
\eq{denk1}{\mathcal{H}=-\summ{<i,j>}{}{\paran{J_{ij}^x s_i^xs_j^x+J_{ij}^y s_i^ys_j^y+J_{ij}^z s_i^zs_j^z}}-\summ{i}{}{H_is_i^z}}
where $s_i^x,s_i^y$ and  $s_i^z$ denote the Pauli spin operators at a site $i$. $J_{ij}^x,J_{ij}^y$ and $J_{ij}^z$ stand for the anisotropy in the exchange interactions between the nearest
neighbor spins located at sites $i$ and $j$ and $H_i$ is the longitudinal magnetic field at a site $i$. The first sum is carried over the nearest neighbors of the lattice, while the second one is over all the lattice sites. The exchange interaction components $(J_{ij}^x,J_{ij}^y,J_{ij}^z)$ between the spins on the sites $i$ and $j$ take the values according to the positions of the nearest neighbor spins. The two surfaces of the film have the intralayer coupling components $(J_{1}^x,J_{1}^y,J_{1}^z)$. The interlayer coupling between the surface and its adjacent layer (i.e. layers $1,2$ and $L-1,L$) is denoted by $(J_{2}^x,J_{2}^y,J_{2}^z)$. For the rest of the layers, the interlayer and the intralayer couplings are assumed as $(J_{2}^x,J_{2}^y,J_{2}^z)$.

We use the two spin cluster approximation as an EFT formulation, namely EFT-2 formulation\cite{ref48}. In this approximation, we choose
two spins (namely $s_1$ and $s_2$) in each layer and treat interactions exactly in this two spin cluster. In order to avoid some mathematical difficulties,
we replace the perimeter spins of the two spin cluster by Ising spins (axial approximation) \cite{ref49}. After all, by using the differential operator technique
and decoupling approximation (DA) \cite{ref50}, we can get an expression for the magnetization per spin, i.e. $m=\sandd{\frac{1}{2}\paran{s_1^z+s_2^z}}$. In the thin film geometry
number of $L$ different representative magnetizations for the system by following the procedure given in Ref. \cite{ref40} can be given as,
\eq{denk2}{\begin{array}{lcl}
m_1&=&\sandd{\Theta_{1,1}^3\Theta_{2,2}}f_1\paran{x,y,H_1,H_2}|_{x=0,y=0}\\
m_k&=&\sandd{\Theta_{2,k-1}\Theta_{2,k}^3\Theta_{2,k+1}}f_2\paran{x,y,H_1,H_2}|_{x=0,y=0},k=2,3,\ldots,L-1\\
m_L&=&\sandd{\Theta_{2,L-1}\Theta_{1,L}^3}f_1\paran{x,y,H_1,H_2}|_{x=0,y=0}.\\
\end{array}} Here $m_i,(i=1,2,\ldots, L)$ denotes the magnetization of the $i^{th}$ layer. The operators in Eq. \re{denk2} are defined via
\eq{denk3}{
\Theta_{k,l}=\left[A_{kx}+m_lB_{kx}\right]\left[A_{ky}+m_lB_{ky}\right]
} where
\eq{denk4}{\begin{array}{lcl}
A_{km}&=&\cosh{\paran{J_k^z\nabla_m}}\\
B_{km}&=&\sinh{\paran{J_k^z\nabla_m}},\quad k=1,2; m=x,y.
\end{array}
}
The functions in Eq. \re{denk2} are given by
\eq{denk5}{f_n\paran{x,y,H_1,H_2}=\frac{x+y+H_1+H_2}{X_0^{(n)}}\frac{\sinh{\paran{\beta X_0^{(n)}}}}{\cosh{\paran{\beta X_0^{(n)}}}+\exp{\paran{-2\beta J_n^z}}\cosh{\paran{\beta Y_0^{(n)}}}}} where
\eq{denk6}{\begin{array}{lcl}
X_0^{(n)}&=&\left[\paran{J_n^x-J_n^y}^2+(x+y+H_1+H_2)^2\right]^{1/2}\\
Y_0^{(n)}&=&\left[\paran{J_n^x+J_n^y}^2+(x-y+H_1-H_2)^2\right]^{1/2}\\
\end{array}}with the values $n=1,2$. In Eq. \re{denk5}, we set $\beta=1/(k_B T)$ where $k_B$ is Boltzmann
constant and $T$ is the temperature.

Magnetization expressions given in closed form in Eq. \re{denk2} can be constructed via acting differential operators on related functions. The effect of the exponential
differential operator to an arbitrary  function $F(x)$ is given by
\eq{denk6}{\exp{\paran{a\nabla}}F\paran{x}=F\paran{x+a}} with any
constant  $a$.

With the help of the Binomial expansion, Eq. \re{denk2} can be written in the form
\eq{denk8}{\begin{array}{lcl}
m_1&=&\summ{p=0}{3}{}\summ{q=0}{3}{}\summ{r=0}{1}{}\summ{s=0}{1}{}K_1\paran{p,q,r,s}m_1^{p+q} m_2^{r+s}\\
m_k&=&\summ{p=0}{1}{}\summ{q=0}{1}{}\summ{r=0}{3}{}\summ{s=0}{3}{}\summ{t=0}{1}{}\summ{v=0}{1}{}K_2\paran{p,q,r,s,t,v}m_{k-1}^{p+q} m_k^{r+s}m_{k+1}^{t+v}\\
m_L&=&\summ{p=0}{3}{}\summ{q=0}{3}{}\summ{r=0}{1}{}\summ{s=0}{1}{}K_1\paran{p,q,r,s}m_L^{p+q} m_{L-1}^{r+s}\\
\end{array}} where

\eq{denk9}{\begin{array}{lcl}
K_1\paran{p,q,r,s}&=&\komb{3}{p}\komb{3}{q}A_{1x}^{3-p}A_{1y}^{3-q}A_{2x}^{1-r}A_{2y}^{1-s}B_{1x}^{p}B_{1y}^{q}B_{2x}^{r}B_{2y}^{s}f_1\paran{x,y,H_1,H_2}|_{x=0,y=0}\\
K_2\paran{p,q,r,s,t,v}&=&\komb{3}{r}\komb{3}{s}A_{2x}^{5-(p+r+t)}A_{2y}^{4-(q+s+v)}B_{2x}^{p+r+t}B_{2y}^{q+s+v}f_2\paran{x,y,H_1,H_2}|_{x=0,y=0}.\\
\end{array}}
These coefficients can be calculated from definitions given in Eq. \re{denk4} by using Eqs. \re{denk5} and \re{denk6}.

For a given Hamiltonian parameters and temperature, by determining the coefficients  from Eq. \re{denk9} we can obtain a system of coupled non linear equations from Eq. \re{denk8}, and by solving this system we can get the longitudinal magnetizations of each layer ($m_i,i=1,2,\ldots,L$). The total longitudinal magnetization ($m$) can be calculated via
\eq{denk10}{m=\frac{1}{L}\summ{i=1}{L}{m_i}.}

Since all longitudinal magnetizations are close to zero in the vicinity of the second order critical point, we can obtain another coupled  equation system to determine the transition temperature by linearizing the equation system given in  Eq. \re{denk8}, i.e.
\eq{denk11}{\begin{array}{lcl}
m_1&=&\parank{K_1\paran{1,0,0,0}+K_1\paran{0,1,0,0}}m_1+\\
&&\parank{K_1\paran{0,0,1,0}+K_1\paran{0,0,0,1}}m_2\\
m_k&=&\parank{K_3\paran{1,0,0,0,0,0}+K_3\paran{0,1,0,0,0,0}}m_{k-1}+\\
&&\parank{K_3\paran{0,0,1,0,0,0}+K_3\paran{0,0,0,1,0,0}}m_{k}+\\
&&\parank{K_3\paran{0,0,0,0,1,0}+K_3\paran{0,0,0,0,0,1}}m_{k+1}\\
m_L&=&\parank{K_1\paran{1,0,0,0}+K_1\paran{0,1,0,0}}m_L+\\
&&\parank{K_1\paran{0,0,1,0}+K_1\paran{0,0,0,1}}m_{L-1}.\\
\end{array}}

Critical temperature, as well as  critical transverse field can be determined from $\mathbf{\mathrm{det(A)=0}}$ where $A$ is the matrix of coefficients of the linear equation system given in Eq. \re{denk11}.

\section{Results and Discussion}\label{results}

In this section, we treat the anisotropic Heisenberg model on the thin film geometry for several cases, with no magnetic field (i.e. $H_1=H_2=0$ in the formulation). The system has
two different exchange interactions which control the spin-spin interactions between the spins belonging to interior ($J_2^x,J_2^y,J_2^z$) and the surfaces ($J_1^x,J_1^y,J_1^z$) of the film.
According to the relations between the components of these exchange interactions, let us entitle the case
$J_1^x=J_1^y=J_1^z$ and ($J_2^x=J_2^y=J_2^z$) as a thin film with isotropic surfaces (interior), and the case $J_1^x=J_1^y, J_1^z=1$ and ($J_2^x=J_2^y, J_2^z=1$) as a thin film with XXZ type surfaces (interior).  In other words, we group the portions of the film (interior and surfaces)  according to the symmetry properties of the relevant exchange interactions as isotropic or XXZ type.

Let us select the  unit of energy as $J$ ($J>0$) and scale all components of the exchange interaction with $J$,
\eq{denk12}{
r_i^{\nu}=\frac{J_i^{\nu}}{J}, \quad i=1,2, \nu=x,y,z
} here since we deal with the ferromagnetic interactions, defined dimensionless parameters in Eq. \re{denk12} are all positive or zero.  Let the parameter $q_\nu, (\nu=x,y,z)$ controls the ratio of the same components of exchange interactions between the surfaces and other layers,
\eq{denk13}{
q_\nu=\frac{r_1^{\nu}}{r_2^{\nu}}, \quad \nu=x,y,z
} where $r_1^{\nu}$ stands for the scaled (dimensionless) $\nu$ component of the exchange interaction between the spins, which belong to surface, while  $r_2^{\nu}$ represents the
$\nu$ component of the exchange interaction between the remaining spins.

\subsection{Isotropic Model}

In this case, all components of the exchange interactions are equal to each other, i.e.
\eq{denk14}{
r_i^{x}=r_i^{y}=r_i^{z}=r_i, \quad i=1,2
} and choose the unit of energy such that $r_2=1$, i.e. unit of energy in Eq. \re{denk12} is $J=J_2^z(=J_2^x=J_2^y)$ . The film consist of isotropic surfaces on an isotropic interior.

The thermodynamic properties of the system are affected by three parameters, which arethe film thickness ($L$), ratio between the surface intralayer exchange interaction and the other exchange interaction  ($r_1$), and the temperature ($k_BT/J$). Thus the critical temperature of the system depends on the parameters $L$ and $r_1$. In a system with a surface, a typical phenomena occurs and this phenomena shows itself in a thin film geometry. The phase diagrams for different film thickness  ($L$) intersect at a special point which can be denoted by $(k_BT_c^{*}/J,r_1^{*})$ in the ($k_BT_c/J,r_1$) plane.  For the values of $r_1>r_1^{*}$, thin films have higher critical temperature than the corresponding bulk system while the reverse is valid for $r_1<r_1^{*}$.
This means that magnetically disordered surface can coexist with a magnetically ordered bulk phase for the values of $r_1$ that provide $r_1<r_1^{*}$ while for the values $r_1>r_1^{*}$, surface can reach the magnetically ordered phase before the bulk.

For the value of $r_1=r_1^{*}$, thin film with thickness $L$ has the same critical values as that of the bulk, independent of the $L$. The value of $r_1$ also changes  the relation between the film thickness and  the critical temperature of the film. The thicker films have higher critical values for $r_1<r_1^{*}$ while thicker films have lower critical values for $r_1>r_1^{*}$ than the thinner ones.

The phase diagrams for the isotropic Heisenberg model in a thin film geometry in a $(k_BT_c/J,r_1)$ plane can be seen in Fig. \re{sek1}. Above mentioned typical behavior can be
seen in this figure. For the special point coordinate  ($k_BT_c^{*}/J,r_1^{*}$), EFT-2 formulation gives ($4.8910,1.3454$), where the first one is just the
critical temperature of the corresponding bulk system (the system with simple cubic lattice) in the same model\cite{ref48}. This value can
be compared with the same system within the high temperature series expansion method ($4.395,1.325$)  \cite{ref35} and
the value of the classical counterpart of the same system within the same formulation ($5.030,1.332$)\cite{ref40_ek}. It is a known fact that the classical system has slightly higher critical
temperature. Also, comparison of the values shows that EFT-2 formulation gives higher value for the $k_BT_c^{*}/J$ and $r_1^{*}$, according to the high temperature series expansion method.


One of the efficient way of describing the behavior of the order parameter with temperature and $r_1$ is to treat the equally valued magnetization curves in the   $(k_BT_c/J,r_1)$ plane. In   Fig. \re{sek2}, we depict this for the film thickness values of $L=3$ and $L=10$. As seen in Figs. \re{sek2} (a) and (b) the variation of the magnetization with these two parameters is qualitatively the same. But the curves become almost parallel lines with the $r_1$ axes when film thickness increases. This means that, magnetization value gets more and more affected by $r_1$, when the film thickness decreases.

The relation between the critical temperature of the film and the corresponding bulk system shows itself also in the
relation between the magnetization of the surface layer and middle layer. In other words, chosen $r_1$ also determines the relation between the magnetization of the surface layer and middle layer for any temperature, which is below the critical temperature. As an example of this situation, the variation of the magnetization of the surface and middle layer as well as the film with temperature can be seen in Fig. \re{sek3} for $L=7$, (a) $r_1=0.5<r_1^{*}$ and  (b) $r_1=2.0>r_1^{*}$. As seen in Fig. \re{sek3} (a), for the values of
$r_1<r_1^{*}$, surface layer has lower magnetization than the middle layer and the situation gets reverse when $r_1>r_1^{*}$ (Fig. \re{sek3} (b)). Of course this situation is valid for the temperatures that are not so close to zero. Even though $r_1=0.0$, the ground state of the surface layer will be ordered and the value is $1.0$. In the absence of the interaction between the spins
on the surface layer ($r_1=0.0$), the surface is completely ordered due to the interaction of the surface with the neighbor layer. The order of the inner layers dictates the  order of the
surface layer in the isotropic interaction case.

\begin{figure}[h]\begin{center}
\epsfig{file=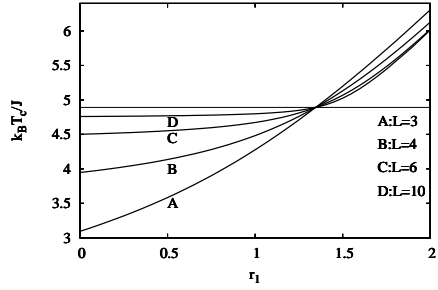, width=6cm}
\end{center}
\caption{Variation of the critical temperature with $r_1$ for some selected film thickness values of $L=3,4,6,10$ and $r_2=1.0$. The horizontal line represents the critical temperature of the isotropic Heisenberg model with the simple cubic lattice structure. } \label{sek1}\end{figure}

\begin{figure}[h]\begin{center}
\epsfig{file=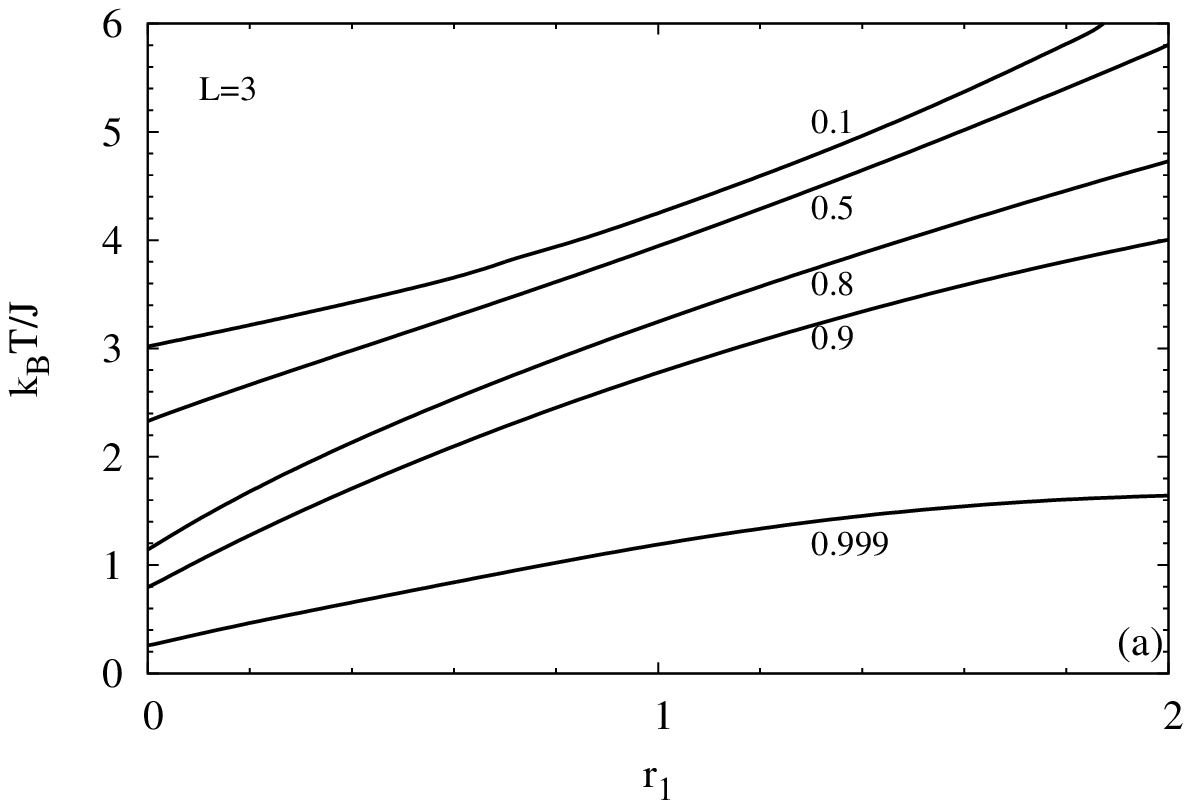, width=6cm}
\epsfig{file=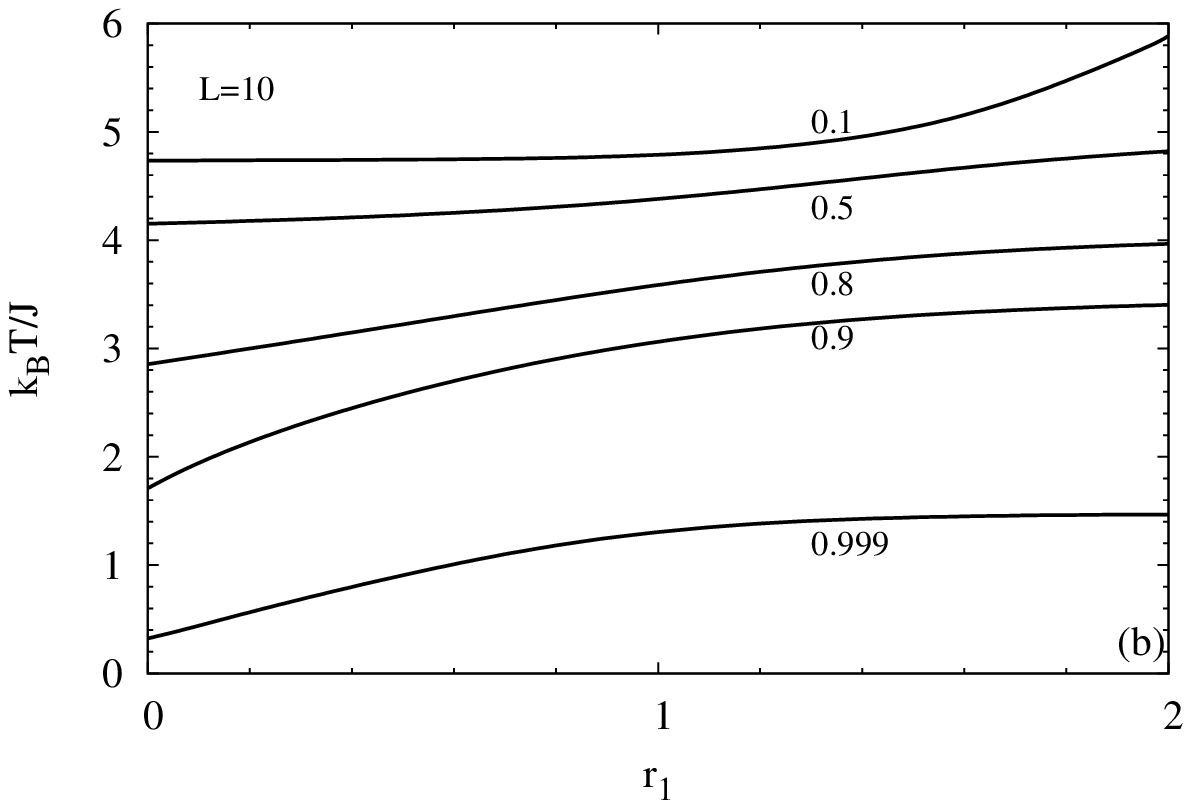, width=6cm}
\end{center}
\caption{Equally valued magnetization curves of the isotropic Heisenberg model in a ($k_BT/J-r_1$) plane with the film thickness values (a) $L=3$ and (b) $L=10$ and $r_2=1.0$. The values of the magnetizations have been underlined near the related curve.
} \label{sek2}\end{figure}

\begin{figure}[h]\begin{center}
\epsfig{file=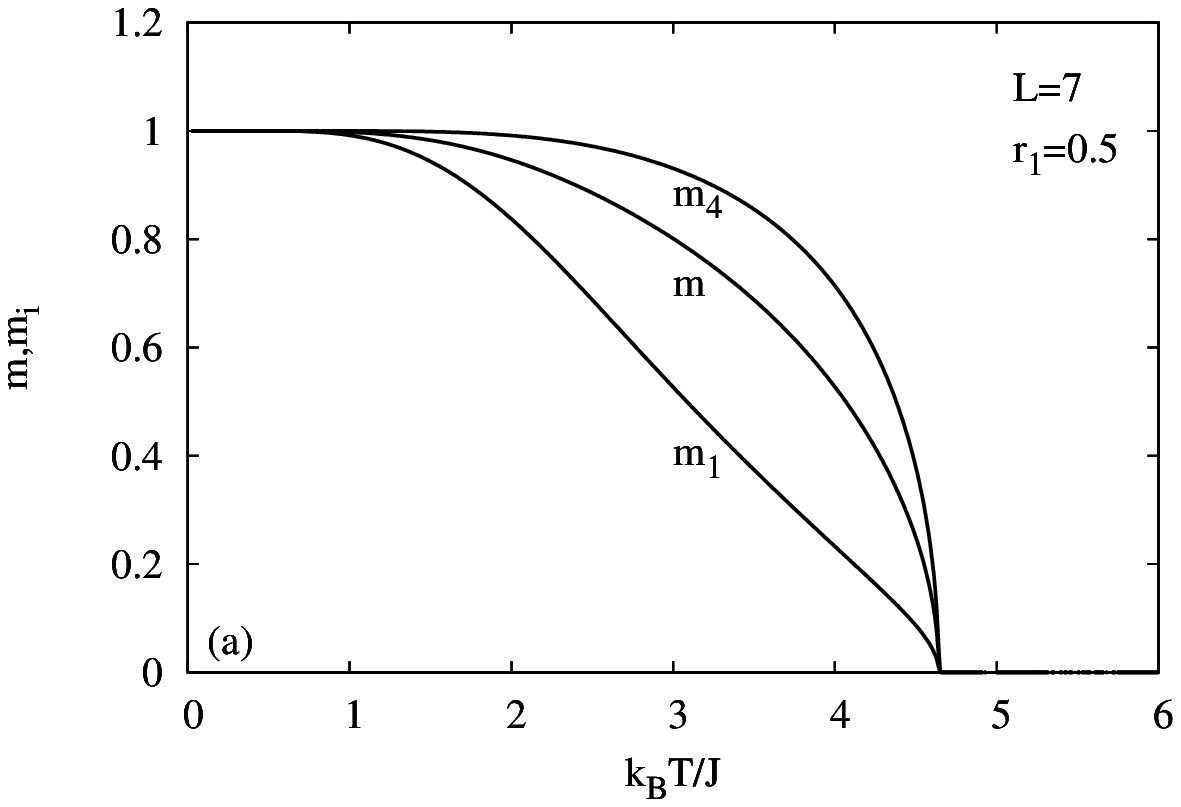, width=6cm}
\epsfig{file=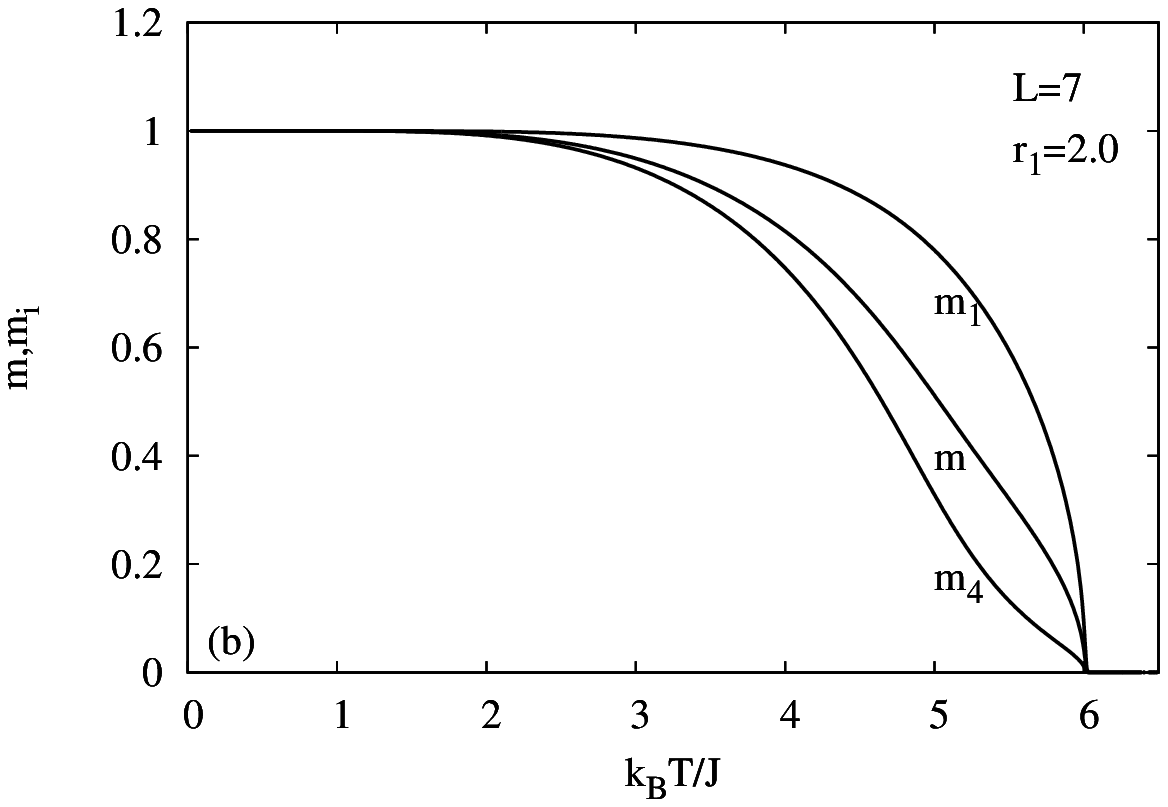, width=6cm}
\end{center}
\caption{Variation of the magnetization of the surface layer ($m_1$), inner layer ($m_4$) and total ($m$) magnetization  with temperature
for $L=7$ and selected values of (a) $r_1=0.5$ and (b) $r_1=2.0$ and $r_2=1.0$. 	
} \label{sek3}\end{figure}

\subsection{Anisotropic Model}

In this case the anisotropies in the exchange interactions are related to each other as
\eq{denk15}{
r_i^{x}=r_i^{y}=r_i, \quad r_i^{z}=1, \quad i=1,2
}
In order to investigate the effect of the anisotropy in the exchange interaction on the critical properties of the system, we choose for the $z$ components of the exchange interactions as $r_1^{z}=r_2^{z}=1$. This means that we choose the unit of energy in Eq. \re{denk12} as $J=J_2^z(=J_1^z)$. From  Eq. \re{denk13} and according to Eq. \re{denk15},  $q_x=q_y$ and let $q=q_x=q_y$. Thus, the parameter $q$ controls the ratio of the anisotropies between the surfaces and other layers. We can treat this case in two sub-cases whether the interior of the film is isotropic or not.

\subsubsection{XXZ-type surfaces on an isotropic interior }

In this case, the inner layers have isotropic exchange interaction both along the interlayer and intralayer and, in order to focus on the effect of the
anisotropy in the exchange interaction which comes from the surfaces, let us choose $r_2^x=r_2^y=1$. In other words, the film consist of a body which has isotropically interacting spins,
coated with surfaces containing intralayer spin-spin interactions which are XXZ-type symmetric.
Thus, the anisotropy in the exchange interaction of the system comes only from the surfaces of the film. In order to see the effect of the anisotropy in the
exchange interaction which comes from the surfaces, we depict the phase diagrams in a $(k_BT_c/J,q)$ plane in Fig. \re{sek4}.
At first sight, we can see from Fig. \re{sek4} that, rising anisotropy in the exchange interaction decreases the critical temperature of the film. Most affected films
are the thinner ones (compare the curves labeled by A and D in Fig. \re{sek4}). Since the surface and the inner layers are interacting via the exchange
interaction $J_2$, the effect of anisotropy can diffuse the inner layers of the film and effect the critical temperature of the system. This propagation becomes easier for the thinner films in comparison with the thicker ones, as seen in Fig. \re{sek4}. Also, as shown in Fig. \re{sek4}, all critical temperatures lie below the
critical temperature of the corresponding bulk system of simple cubic geometry and it is shown as a thinner horizontal line in the diagram.

\begin{figure}[h]\begin{center}
\epsfig{file=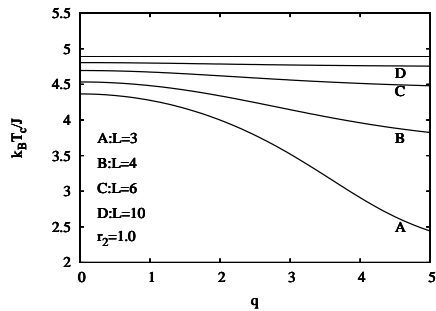, width=6cm}
\end{center}
\caption{Variation of the critical temperature with $q$ for some selected film thickness values of $L=3,4,6,10$. The horizontal thinner line represents the critical temperature of the isotropic Heisenberg model for a simple cubic lattice.
} \label{sek4}\end{figure}

\begin{figure}[h]\begin{center}
\epsfig{file=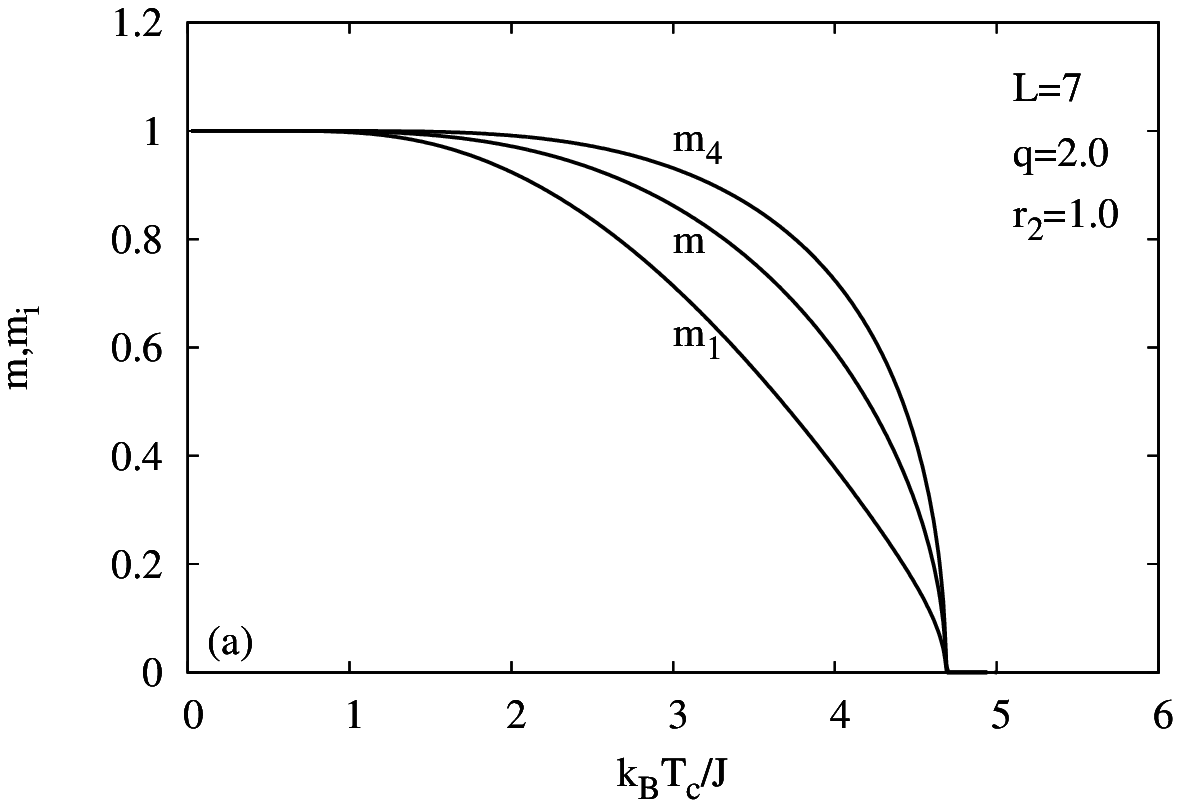, width=6cm}
\epsfig{file=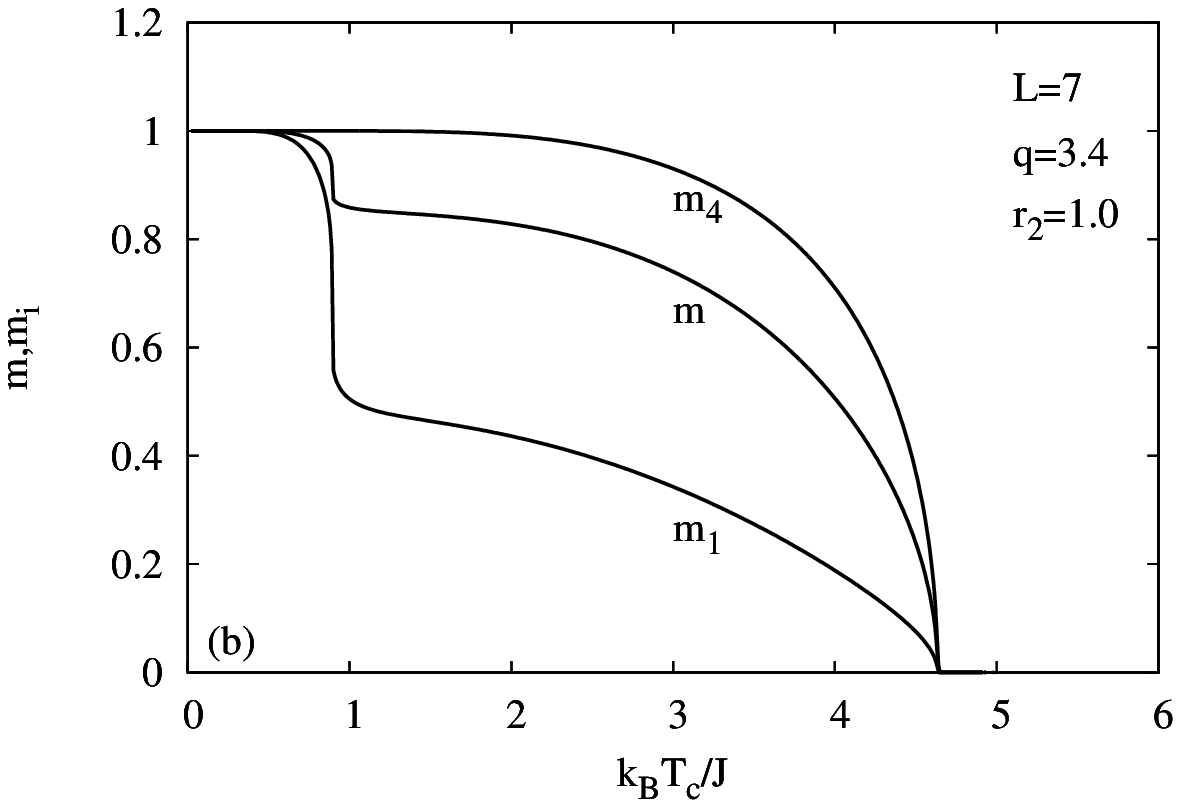, width=6cm}
\epsfig{file=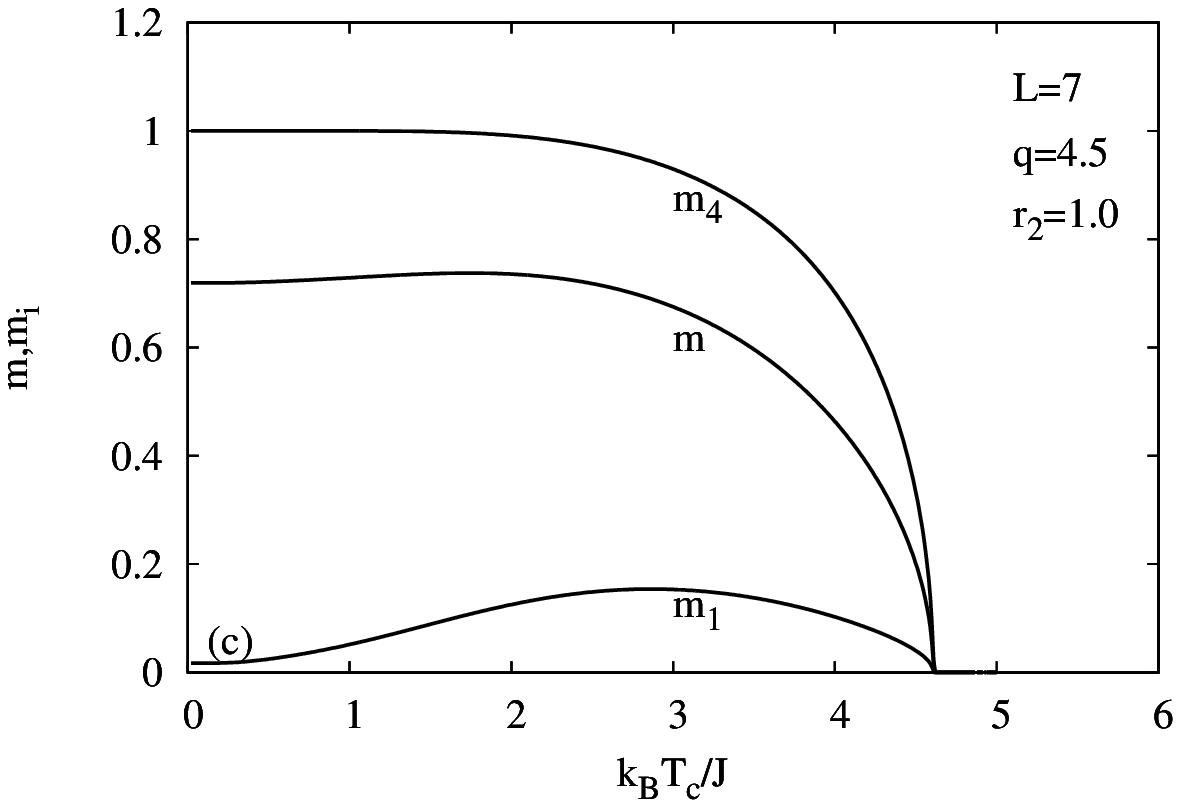, width=6cm}
\epsfig{file=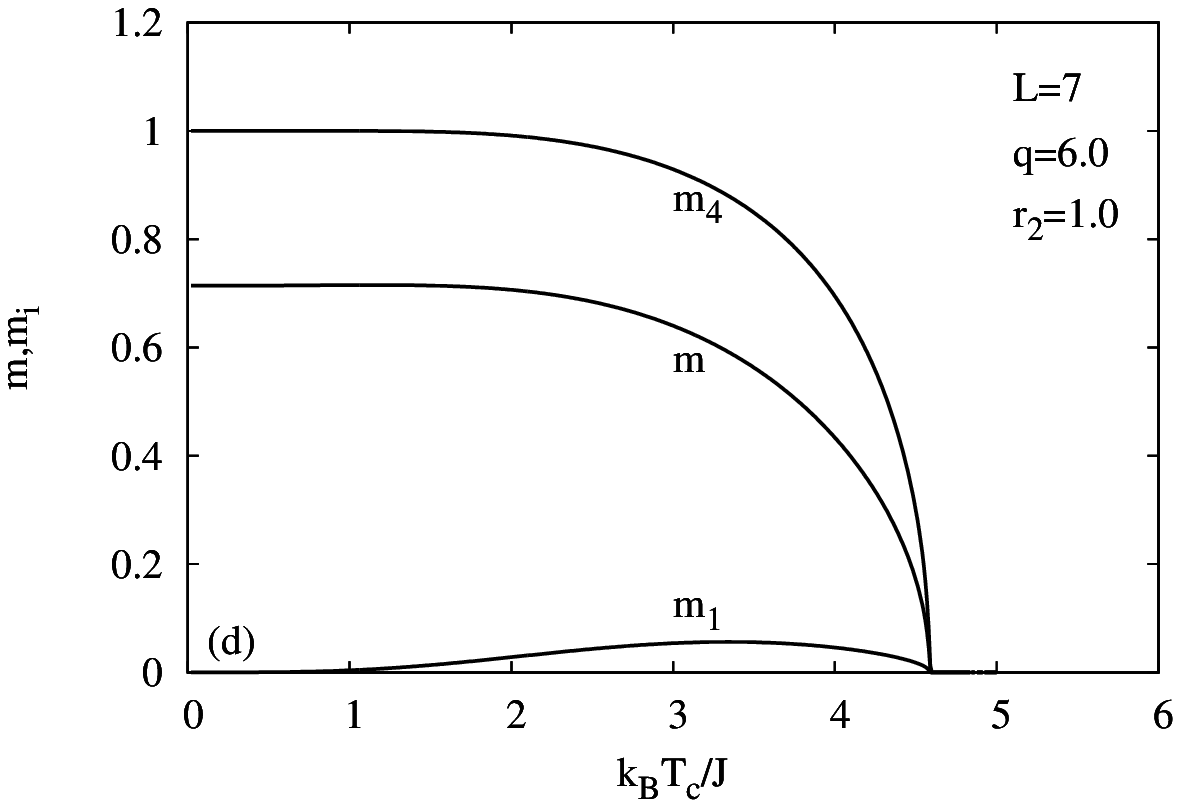, width=6cm}
\end{center}
\caption{Variation of the magnetization of the surface layer ($m_1$), inner layer ($m_4$) and total ($m$) magnetization  with the temperature
for $L=7$ and selected values of (a) $q=2.0$, (b) $q=3.4$, (c) $q=4.5$ and (d) $q=6.0$.   	
} \label{sek5}\end{figure}

When the surface anisotropy is higher than that of the interior of the film, surface magnetization has a lower value than that of the interior. This situation can be seen in  Fig. \re{sek5} for a film that consist of XXZ-type surfaces on an isotropic interior, with $L=7$. In Fig. \re{sek5}, $m$ stands for the total magnetization of the film, while $m_1$ denotes the magnetization of the
surface layer and $m_4$ denotes the magnetization of the middle layer. As seen in Fig. \re{sek5}, rising $q$ (which means rising anisotropy in the exchange interaction of the surfaces) depresses the surface layer magnetization to zero. But as seen in Fig. \re{sek5} (d), while the surface has zero magnetization at lower temperatures, rising temperature gives rise to non-zero magnetization value for the surface layer. Thermal agitations can give rise to surface magnetization which is different from zero at high anisotropy values in the exchange interaction of the surfaces.

\subsubsection{XXZ-type surfaces on a XXZ-type interior}

In this case, both of the exchange interactions of the system are anisotropic as XXZ-type. Here Eq. \re{denk15} is valid but in contrast to previous subsection, $r_2$ is different from unity. The problem of the behavior of the order parameter has four parameters, namely $r_1,r_2,L$ and $k_BT/J$. Thus, the critical temperature of the system depends on the number of layers of the system ($L$), anisotropy in the exchange interaction of the surface ($r_1$) and the anisotropy in the exchange interaction of the inner layers ($r_2$).

The phase diagrams in a $(k_BT_c/J,r_2)$ plane can be seen in Fig. \re{sek6} for some selected values of $q=r_1/r_2$. The thinner lines in each figure corresponds to the
variation of the critical temperature of the corresponding bulk system which has spin-spin interaction ($r_2$). The symmetry of this interaction is given in Eq. \re{denk15}.  Fig. \re{sek6}(a)
corresponds to the system that has Ising type surfaces on a XXZ-type interior,
while Fig. \re{sek6}(c) corresponds to the system with XXZ-type surfaces on a XXZ-type interior with the same magnitudes of anisotropy. The other two panels correspond to the system of XXZ-type surfaces on a XXZ-type interior, but in Fig. \re{sek6}(b), the  surface anisotropy value is lower than that of the interior  ($q=0.5$), whereas in Fig. \re{sek6}(d), the surface anisotropy value is greater than that of the interior  ($q=3.0$).

First, rising anisotropy in the exchange interaction of the interior
($r_2$) decreases the critical temperature of the system for any $q$, as expected. The important point is that the special point appears at the lower values of $q$ and it disappears while $q$ rises. Let us denote the coordinates of this special point as $(k_BT_c^*/J,r_2^*)$. The evolution of the special point coordinates with $q$ can be seen in Fig. \re{sek7}. As seen in Fig. \re{sek7}, special point  disappears just after the value of  $q=0.83$.  Thus, if the ratio of the surface anisotropy to the interior anisotropy value is  greater than $q=0.83$, then thinner films have lower critical temperature for all values of the $q$. For the values that provide $q<0.83$ special point can appear (as seen in Fig. \re{sek6} (a) and (b)) and in this case, for the anisotropy values that provide $r_2>r_2^*$ with $q<0.83$,  it is possible to observe thinner films with higher critical temperature in comparison with that of the thicker ones. This special point moves on the lines can be seen in Fig. \re{sek7} as $q$ rises.

We can see  from Figs. \re{sek6}  (c) and (d) that, if the special point is not present in the system then all the phase diagrams lie below the phase diagram of the bulk system (denoted by thinner lines). Besides, after some value of anisotropy ($q$), the phase diagrams of the thinner films depress to zero, i.e. all transitions are of the second  order (e.g. the curves labeled by A and B in \re{sek6} (d)). This is interesting since at a certain value of $q$, when the thickness of the film rises, second order transitions turn into first order transitions (compare the curves labeled by B and C in \re{sek6} (d)).

We again note that, some of the phase diagrams in Fig. \re{sek6} depress to zero whereas some of them do not. Since the formulation can give only second order transition temperatures, the phase diagrams that does not depress to zero have the portions of first order transitions which we can not calculate within this formulation.

\begin{figure}[h]\begin{center}
\epsfig{file=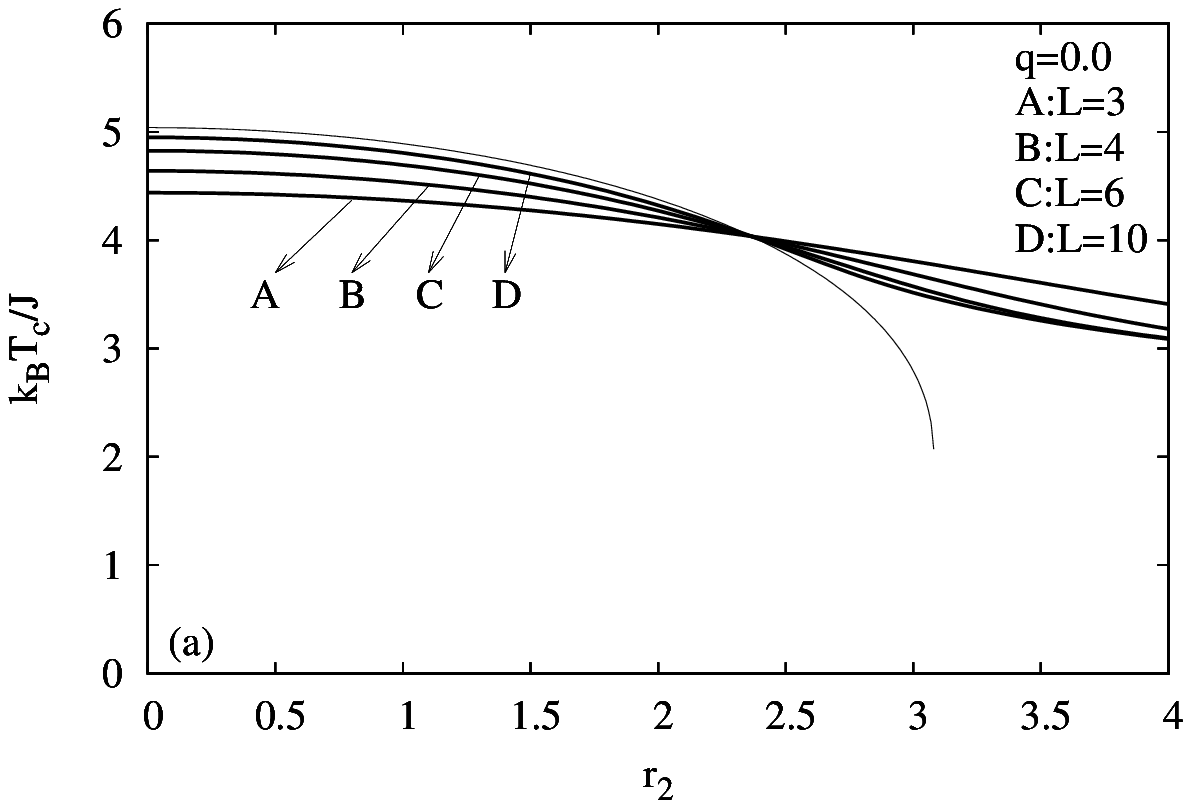, width=6cm}
\epsfig{file=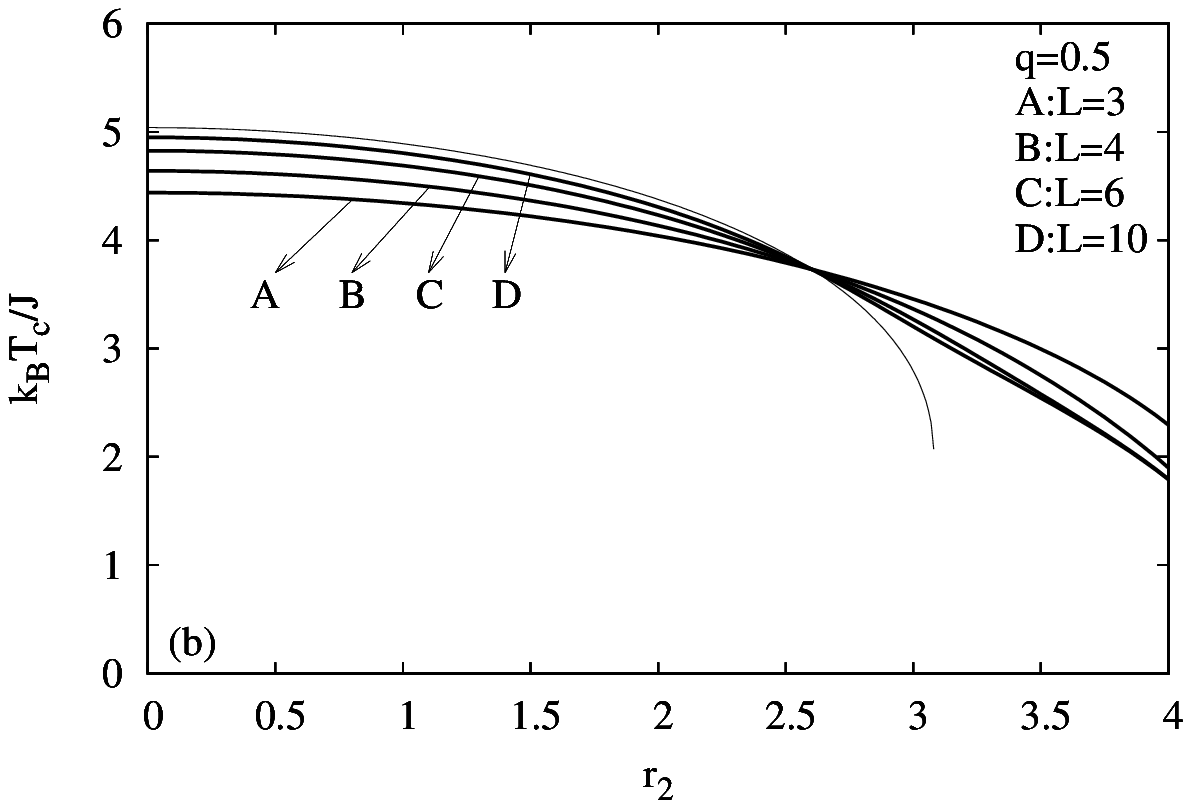, width=6cm}
\epsfig{file=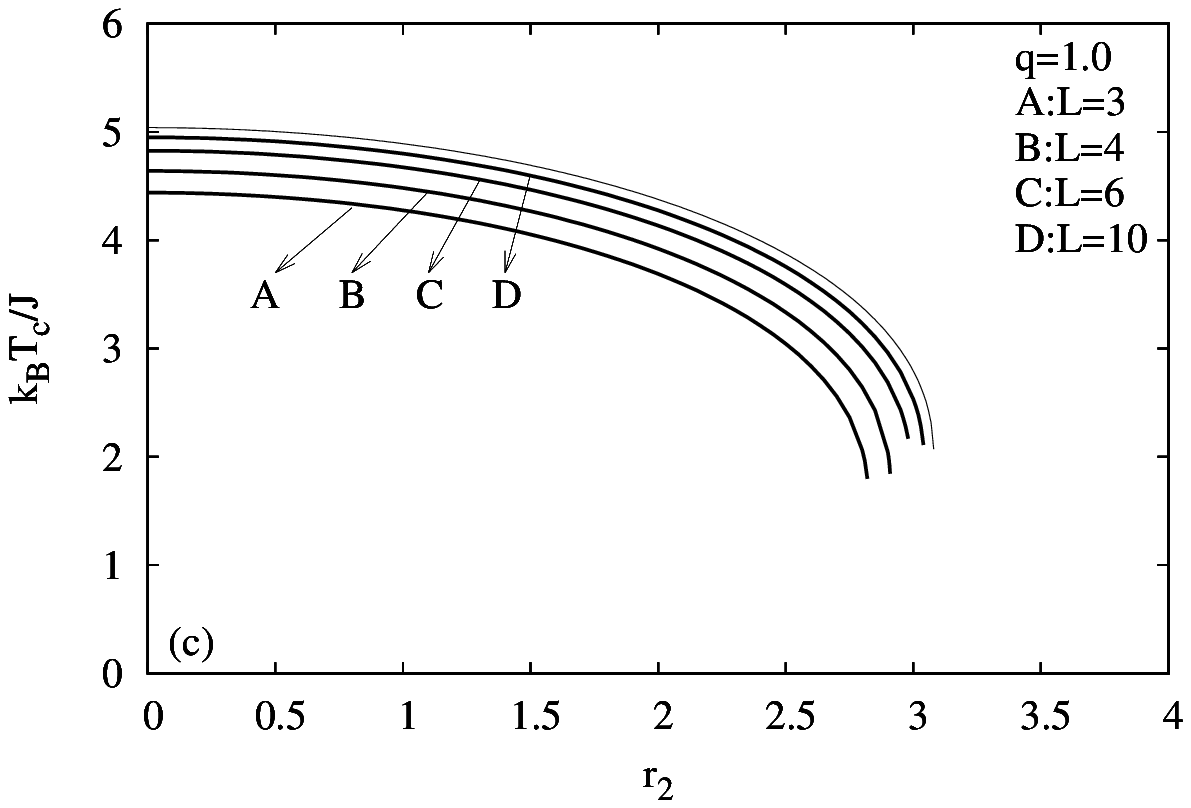, width=6cm}
\epsfig{file=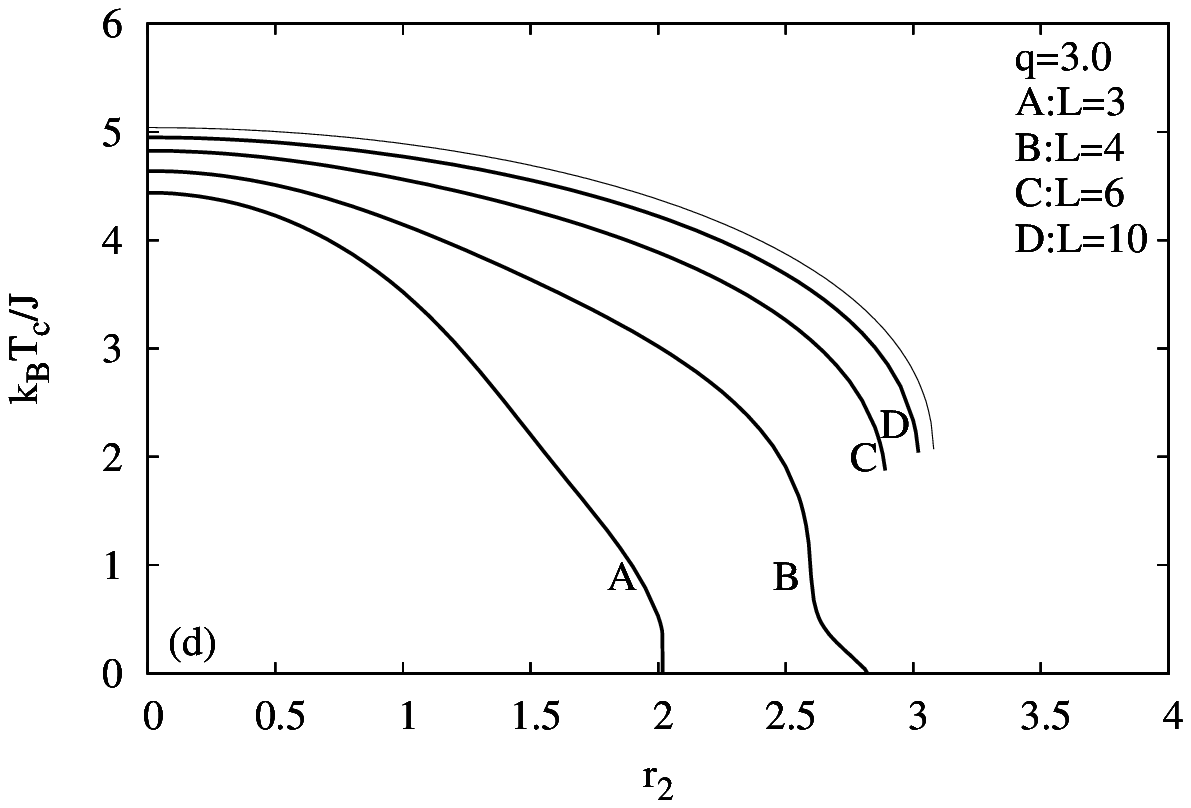, width=6cm}
\end{center}
\caption{The phase diagrams of the thin film consisting of XXZ-type surfaces on a XXZ-type interior, in a $(k_BT_c/J,r_2)$ plane for some selected values of $q$ and for the thickness values of $L=3,4,6,10$
} \label{sek6}\end{figure}

\begin{figure}[h]\begin{center}
\epsfig{file=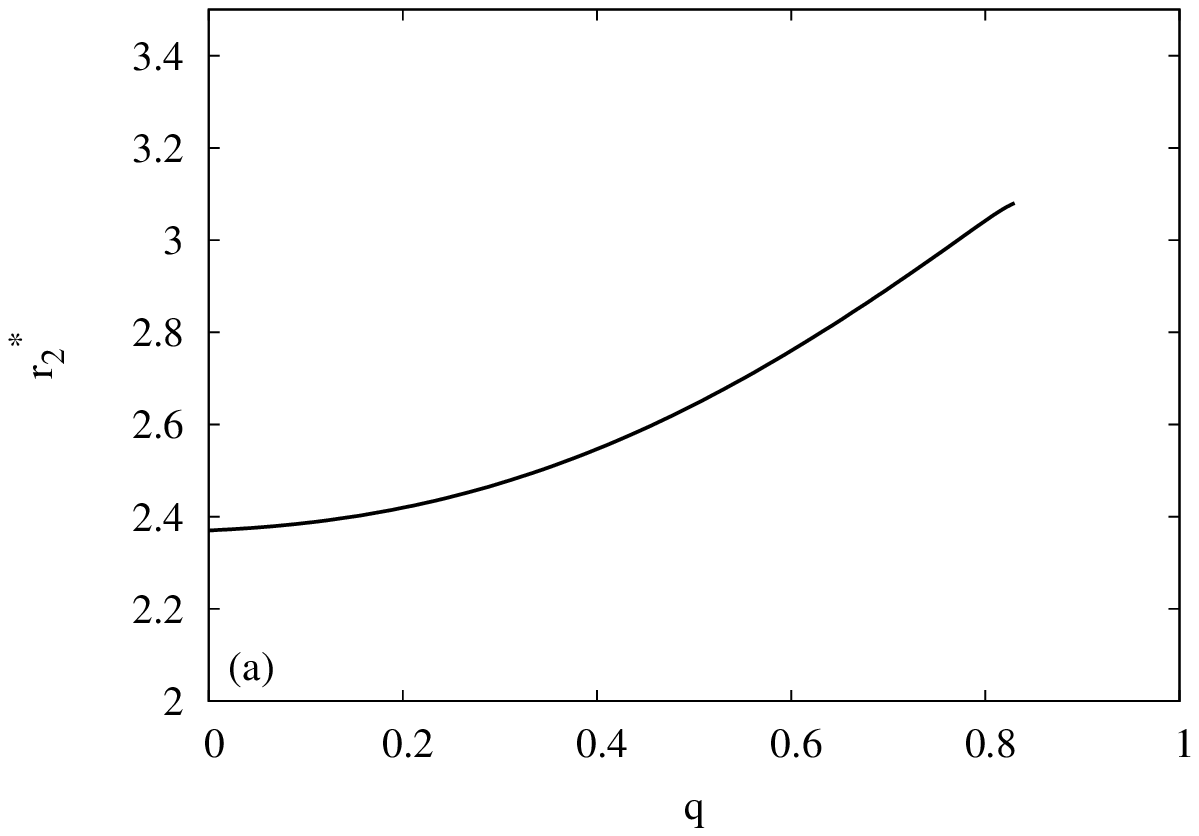, width=6cm}
\epsfig{file=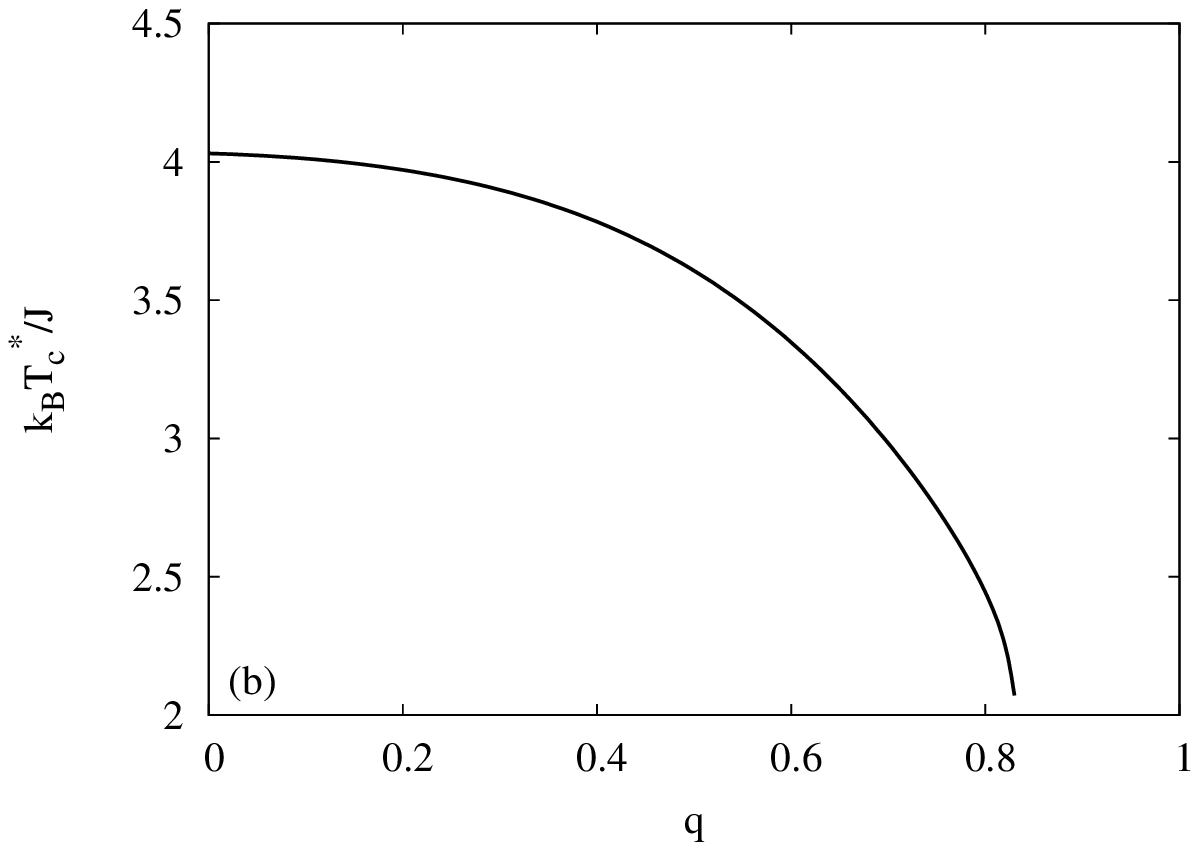, width=6cm}
\end{center}
\caption{Variation of the special point coordinates (a) $r_2^*$ and (b) $k_BT_c^*/J$ with  $q$ for the thin film consisting of XXZ-type surfaces on a XXZ-type interior.
} \label{sek7}\end{figure}

\section{Conclusion}\label{conclusion}

In this work, the effect of the anisotropy in the exchange interaction on the critical properties of the thin films have been investigated.
As a formulation, the differential operator technique and DA within the  EFT-2 formulation has been used.

The system has two different exchange interactions and the problem has been handled by grouping the systems according to whether the surface/interior has anisotropic exchange interaction or not as: isotropic type surfaces on an isotropic type interior, XXZ-type surfaces on an isotropic type interior and XXZ-type surfaces on a XXZ-type interior.

For the critical temperatures of the films, isotropic type surfaces on an isotropic type interior system exhibit special point which has
coordinates
$(k_BT_c^*/J,r_1^*)$= $(4.8910,1.325)$. This special point is related to the relation between the exchange interaction of the surface layer and inner layers. Thus,
the anisotropy ratio between the surface and inner layers  have to effect on this coordinate. This effect has been shown in Fig. \re{sek7}
for the system with XXZ-type surfaces on a XXZ-type interior,
and we have concluded that after a ratio of $q=0.83$, special point does not appear in the system. Besides, in the system with XXZ-type surfaces on the isotropic interior, it has been shown that rising surface anisotropy decreases the critical temperature of the film and most affected films are the thinner ones.

For the thermodynamical properties of the film, in the isotropic case (i.e. isotropic surfaces on an isotropic interior), based on the ratio of the exchange interactions between the surface and interior of the film, surface magnetization can lie below or above the magnetization of the inner layers. This situation is related to whether that ratio is greater than the special point or not. For the values that provide $r_1>r_1^*=1.325$, surface magnetization is greater than the inner layers and vice versa. For the XXZ-type surface with isotropic interior, surface magnetization is lower than the  magnetization of the interior of the film for the values of $q>1.0$, where $q$ relates the  anisotropy in the exchange interaction of the surface with the inner layers. It has been shown that when $q$ increases, the magnetization of the surface layer depresses to zero at low temperatures, but due to rising thermal fluctuation which comes from the rising temperature can create a non-zero magnetization.

We hope that the results  obtained in this work may be beneficial form both theoretical and experimental point of view.

\end{document}